\documentclass[prb,aps,eqsecnum]{revtex4}
\usepackage{graphicx}
\newcommand{\e}{{\rm e}}
\newcommand{\ep}{\varepsilon}

\newcommand{\be}{\begin{equation}}
\newcommand{\ee}{\end{equation}}
\newcommand{\ba}{\begin{eqnarray}}
\newcommand{\ea}{\end{eqnarray}}

\newcommand{\nn}{\nonumber}
\newcommand{\la}{\label} 
\newcommand{\w}{\omega}

\newcommand{\dt}{\Delta\tau}

\begin{document}
\title{Fourth-Order Algorithms for Solving the
Imaginary Time Gross-Pitaevskii Equation in a
Rotating Anisotropic Trap}

\author{Siu A. Chin}
\affiliation{Department of Physics, Texas A\&M University,
College Station, TX 77843, USA}
\author{Eckhard Krotscheck}
\affiliation{Institut f\"ur Theoretische Physik, Johannes Kepler
Universit\"at Linz, A-4040 Linz, Austria}

\begin{abstract}

By implementing the exact density matrix for the rotating 
anisotropic harmonic trap, we derive a class of very fast and accurate
fourth order algorithms for evolving the Gross-Pitaevskii equation
in imaginary time. Such fourth order algorithms are possible only
with the use of {\it forward}, positive time step factorization 
schemes. These fourth order algorithms converge at time-step
sizes an order-of-magnitude larger than conventional second order 
algorithms. Our use of time-dependent factorization schemes 
provides a systematic way of devising algorithms for solving 
this type of nonlinear equations. 

\end{abstract}
\maketitle

\section {Introduction}

The dynamics of a fast rotating	Bose-Einstein condensate (BEC)
has been studied extensively in terms of the 
Gross-Pitaevskii (GP) equation\cite{dalf,fet}. By evolving the GP equation
in imaginary time, it is easy to determine the ground state
properties of the condensate, such as the formation of
vortex-arrays and giant vortices\cite{fet,fetter}. It has been known for 
some time that the first order pseudo-spectral, split-operator 
method\cite{taha} is a very fast way of solving the non-linear 
Schr\"odinger equation. However, first or second order split operator
(SO) methods\cite{jackson,dion} and Crank-Nickolson (CN) algorithms
with\cite{adhi} or without\cite{rupr}
splitting ignore the time-dependence of the non-linear 
potential and converge linearly or quadratically only at very 
small time steps. Bandaruk and Shen\cite{band} have applied 
higher order decomposition schemes with negative coefficients to 
solve the real time non-linear Schr\"odinger equation. 
Due to the difficulty of estimating the non-linear potential 
at intermediate time, they have not demonstrated that their higher 
order algorithms actually converge with accuracy beyond second order. 
In any case, their negative coefficient algorithms cannot be used for 
imaginary time evolution because negative time steps will result in 
an unbounded diffusion kernel\cite{lang4,dmc4,auer,cift03}.

In this work, we derive a class of very accurate fourth order factorization
algorithm for solving the GP equation in imaginary time. These algorithms
are made possible by the confluence of three key ideas: 
1) The density matrix for a rotating anisotropic harmonic oscillator 
can be solved exactly.
2) The time-dependence of the non-linear potential can be systematically
accounted for in factorization algorithms. 
3) Forward, all positive time step 
algorithms\cite{suzfour,chin97,chinchen02,chinchen03,ome02,ome03} 
are now available for solving imaginary time evolution equations.    

In the next section, we show how the density matrix of the harmonic
oscillator can be exactly implemented as an algorithm. 
This obviates the need to
expand in harmonic eigenstates\cite{dion,edwards}. In Section III, 
by exact diagonalization, we generalize 
the result to the case of a rotating anisotropic harmonic trap. 
In Section IV, we describe the
time-dependent form of the fourth-order forward algorithm for solving
the GP equation. In Section V, we compare the convergence of various
algorithms.  We summarize our conclusions in Section VI.

\section {Exact algorithm for the Harmonic oscillator}

Consider the 1-D harmonic oscillator Hamiltonian operator given by 
\be
H=T+V=\frac12 p^2+\frac12 \w^2x^2.
\ee
Its imaginary time propagator
(or density matrix) can be exactly decomposed as
\be
\e^{-\tau (T+V)}=\e^{-\tau C_VV}
   \e^{-\tau C_TT}\e^{-\tau C_VV},
\la{exfact}
\ee
where $C_V$ and $C_T$ are functions of $\tau$ to be
determined. To show this, we simply compare the 
matrix elements on both sides. For the RHS, we have
(ignoring normalization factors)
\be
\langle x^\prime|\e^{-\tau C_VV}
   \e^{-\tau C_T T}\e^{-\tau C_V V}|x\rangle
=
\e^{-\tau C_V\frac12 \w^2{x^\prime}^2}
\e^{-\frac1{2\tau C_T}(x^\prime-x)^2}
\e^{-\tau C_V\frac12 \w^2x^2}.
\la{fact}
\ee 
For the LHS of (\ref{exfact}), the exact density matrix 
element is known\cite{fey}
\ba
\langle x^\prime|\e^{-\tau (T+V)}|x\rangle
&=&\exp\biggl(
-\frac{\w}{2\sinh(\w\tau)}
\Bigl[({x^\prime}^2+x^2)\cosh(\w\tau)-2x^\prime x\Bigr]\biggr)\nn\\
&=&\exp\biggl(
-\frac{\w}{2\sinh(\w\tau)}
\Bigl[({x^\prime}^2+x^2)(\cosh(\w\tau)-1)
+(x^\prime-x)^2\Bigr]\biggr),
\la{exmat}
\ea
where we have expressed
 $-2x^\prime x=(x^\prime-x)^2-{x^\prime}^2-x^2$.
Comparing (\ref{fact}) to (\ref{exmat}) allows us to
identify the coefficient functions as
\be
C_V=\frac{\cosh(\w\tau)-1}{\w\tau\sinh(\w\tau)}
\quad{\rm and}\quad
C_T=\frac{\sinh(\w\tau)}{\w\tau}.
\la{ct}
\ee
In the limit of $\tau\rightarrow 0$, we have
\ba
C_V&=&\frac12-\frac1{24}\w^2\tau^2+\frac1{240}\w^4\tau^4+\cdots,
\la{cvexp}\\
C_T&=&\,1+\,\,\frac16\, \w^2\tau^2 +\frac1{120}\w^4\tau^4+\cdots.
\la{ctexp}
\ea
If we keep only the first term, we have a second order algorithm.
Keeping the first two terms gives a fourth order algorithm,
keeping the first three terms gives a sixth order algorithm, etc..

The exact factorization (\ref{exfact}) is possible
for the harmonic oscillator because its Hamiltonian is quadratic 
and higher order commutators are either zero or simply 
proportional to $T$ or $V$. The harmonic oscillator is characterized
by two key commutators, 
\be
[V,[T,V]]=2\w^2V,
\la{vtv}
\ee
\be 
\,\,\,\,\,\,[T,[V,T]]=2\w^2T.
\la{ttv}
\ee
Because of these two equalities, all higher order commutators 
can be subsumed back to the original operators $T$ and $V$. 
To see how this
exact decomposition comes about, let's begin with the 
simple second order decomposition,
\ba
{\rm e}^{-{1\over 2}\tau V}{\rm e}^{-\tau T}
{\rm e}^{-{1\over 2}\tau V}
&=&\exp\Bigl[-\tau(T+V)
+{1\over{24}}\tau^3([V,[T,V]]-2[T,[V,T]])+O(\tau^5)\,\Bigr],\nn\\
&=&\exp\Bigl[-\tau(T+V)
+{1\over{24}}\tau^3(2\w^2 V-4\w^2 T)+O(\tau^5)\,\Bigr].
\label{sym2}
\ea
Since the error terms are proportional to the original operators,
they can be symmetrically moved back to the LHS to yield,
\be
{\rm e}^{-\tau(\frac12-\frac1{24}\w^2\tau^2) V}
{\rm e}^{-\tau(1+\frac16 \w^2\tau^2) T}
{\rm e}^{-\tau(\frac12-\frac1{24}\w^2\tau^2) V}
=\e^{-\tau(T+V)+O(\tau^5)}.
\label{sym22}
\ee
The decomposition of the LHS is then correct to fourth order. The
coefficients agree with the expansion (\ref{cvexp})
and (\ref{ctexp}). This example makes it clear that
the exact expansion only depends on the abstract
commutator relations (\ref{vtv}) and (\ref{ttv}), and is
independent of the specific representation of
the 1-D harmonic oscillator. Also, if
we exchange the operators $T \leftrightarrow V$,
the coefficients are unchanged. Thus we can also
factorize exactly via
\be
\e^{-\tau (T+V)}=\e^{-\tau C_VT}
   \e^{-\tau C_TV}\e^{-\tau C_VT}.
\la{exfactd}
\ee

For real time propagation, we only need to set
$\tau=it$ to get the corresponding coefficients,
\be
C_V=\frac{1-\cos(\w t)}{\w t\sin(\w t)}
\quad{\rm and}\quad
C_T=\frac{\sin(\w t)}{\w t}.
\la{ctr}
\ee

For either real or imaginary time evolution, one iterates the
discretized wave function forward in time via
\be
|\psi(\tau+\Delta\tau)\rangle=\e^{-\Delta\tau (T+V)}|\psi(\tau)\rangle
\ee
If the exact density matrix (\ref{exmat}) were used directly
in coordinate space, that would incur a slow, $N\times N$
matrix multiplication of the Gaussian kernel
$\e^{-\frac1{2\tau C_T}(x^\prime-x)^2}$. 
The advantage of the factorized form is that this matrix 
multiplication can be avoided by going to k-space via FFT
and multiplying the k-space wave function point-by-point by
$\e^{-\tau C_T\frac12 k^2}$. This is then an order $N\ln_2\,N$ operation,
much faster than the $N\times N$ coordinate space matrix multiplication.

\section {Exact algorithm for a rotating anisotropic harmonic trap}

Consider now the case of an rotating anisotropic harmonic potential
with Hamiltonian
\be
H=\frac12(p_x^2+p_y^2)
+\frac12{\tilde\omega}^2_x x^2
+\frac12{\tilde\omega}^2_y y^2
-\tilde\Omega\left(xp_y
         -yp_x\right).
\la{hmaghar}
\ee
This is a well-studied problem in nuclear physics\cite{ring}.
Its diagonalization is greatly simplified\cite{ok} if we characterize
the anisotropy via the deformation parameter $\delta$
\be
{\tilde\omega}^2_x=(1+\delta)\omega^2_0,\quad
{\tilde\omega}^2_y=(1-\delta)\omega^2_0,
\ee
measure lengths in units of the oscillator length
$l=1/\sqrt{\omega_0}$, and express $H$ and $\tilde\Omega$ 
in units of $\omega_0$. The resulting dimensionless
Hamiltonian is then
\be
H
=\frac12(p_x^2+p_y^2)
+\frac12(1+\delta)x^2
+\frac12(1-\delta)y^2
-\Omega\left(xp_y
         -yp_x\right),
\la{hroto}
\ee
where $\Omega=\tilde\Omega/\omega_0$.
To diagonalize this Hamiltonian, we introduce
two new sets of canonical variables,
\be
Q_1=\alpha_1(c x-s p_y),\quad
P_1=\frac1{\alpha_1}(c p_x+s y),
\ee 
\be
Q_2=\alpha_2(c y-s p_x),\quad
P_2=\frac1{\alpha_2}(c p_y+s x),
\ee
where $\alpha_i$ are normalization constants,
and
$c=\cos(\phi)$,
$s=\sin(\phi)$. 
One can check that the canonical commutator relations
are indeed satisfied,
\be
\left[Q_i,P_j\right]=\mathrm{i}\delta_{ij}\, .
\nn
\ee
In terms of $\{Q_i,P_i\}$, 
because of the way we have
parameterized the anisotropy and expressed everything
in terms of $\w_0$,
the coefficients of {\it both} $P_2Q_1$ and $P_1Q_2$ can
be made to vanish with a single condition:
\be
\tan(2\phi)=\frac{2\Omega}\delta.
\la{angle}
\ee
Using $\alpha_i$ to normalize the $P_i^2$ terms
with unit coefficient,
the resulting Hamiltonian can be written as
\be
H=T_1+V_1+T_2+V_2=
\frac12 P_1^2+\frac12 \Omega_1^2Q_1^2  
+\frac12 P_2^2+\frac12 \Omega_2^2Q_2^2,
\la{hrot}
\ee 
where 
\ba
\alpha_1^{-2}
&=&1-\frac\delta{2}+\frac12 \sqrt{\delta^2+4\Omega^2}\,,\nn\\
\alpha_2^{-2}
&=&1+\frac\delta{2}-\frac12 \sqrt{\delta^2+4\Omega^2}\,,
\ea
with  
\ba
\Omega_1^{2}
&=&1+\Omega^2+\sqrt{\delta^2+4\Omega^2}\,, \nn\\
\Omega_2^{2}
&=&1+\Omega^2-\sqrt{\delta^2+4\Omega^2}\,.
\ea
Also, from (\ref{angle}), we have 
\ba
2 s^2&=&1-\frac\delta{\sqrt{\delta^2+4\Omega^2}}\,,\nn\\
\quad
2 c^2&=&1+\frac\delta{\sqrt{\delta^2+4\Omega^2}}\,.
\la{angles}
\ea
At $\Omega=0$, the phase angle $\phi=0$.
As $\Omega$ increases, the phase angle approaches
$45^\circ$ asymptotically. Thus $s$ and $c$ 
in (\ref{angles}) are both positive. However, as $\Omega$ increases,
$\Omega^2_2$ crosses zero and becomes negative
at $\Omega=\sqrt{1-\delta}$. At this critical rotation rate,
the Coriolis force overcomes the weaker
harmonic potential in the y-direction and the 
anisotropic harmonic oscillator is unstable.
$\Omega^2_2$ emerges positive
again when $\alpha_2^{-2}$ crosses zero and turn
negative at $\Omega^2=1+\delta$.
Thus $\Omega^2_2$
is negative over the interval
$1-\delta\le \Omega^2\le 1+\delta$. 
This is an instability
of the rotating harmonic oscillator, not necessary that of
the Gross-Pitaevskii equation. We will come back to this
point in Section VI. Note also that for $\delta=0$, the algorithm
is stable up to $\Omega=1$.   

Eq. (\ref{hrot})
consists of two independent harmonic oscillator with different 
frequency. The two exact algorithms must be applied in sequence.
However, since $T_1$ and $V_2$ only depend on $p_x$ and $y$,
they should be placed next to each other so that both can be evaluated in
the same mixed representation described below. Similarly,
$T_2$ and $V_1$ only depend on $x$ and $p_y$.
We therefore use the following factorization for each algorithm,
\be
\e^{-\tau (T_1+V_1)}=
\e^{-\tau C_V(1)T_1}
\e^{-\tau C_T(1)V_1}
\e^{-\tau C_V(1)T_1},
\ee
\be
\e^{-\tau (T_2+V_2)}=
\e^{-\tau C_V(2)V_2}
\e^{-\tau C_T(2)T_2}
\e^{-\tau C_V(2)V_2},
\ee
and interlaced them as follow:
\be
\e^{-\tau (T_1+V_1+T_2+V_2)}=
\e^{-\tau C_V(1)T_1-\tau C_V(2)V_2}
\e^{-\tau C_T(1)V_1-\tau C_T(2)T_2}
\e^{-\tau C_V(1)T_1-\tau C_V(2)V_2}.
\la{exfact2}
\ee
Here we use the shorthand notations $C_V(1)=C_V(\Omega_1)$,
$C_T(2)=C_T(\Omega_2)$, etc..
To implement (\ref{exfact2}), let
\be
\psi(x,y)=\frac{1}{\sqrt{2\pi}}\int dp_x\,\psi(p_x,y)\,\e^{\mathrm{i}p_x x},
\ee
\be
\psi(p_x,y)=\frac{1}{\sqrt{2\pi}}\int dx\, \psi(x,y)\,
\e^{-\mathrm{i}p_x x},\nn\\
\ee
and
\ba
\psi(x,p_y)&=&\frac{1}{\sqrt{2\pi}}\int dy\, \psi(x,y)\,
\e^{-\mathrm{i}p_y y},\nn\\
&=&\frac{1}{2\pi}\int dy\, dp_x\, \psi(p_x,y)\,\e^{\mathrm{i}p_x x-
\mathrm{i}p_y y}.
\ea
The operators $T_1$ and $V_2$ are diagonal in the representation $\psi(p_x,y)$
and $T_2$ and $V_1$ are diagonal in the representation $\psi(x,p_y)$. In
practice, $\psi(x,y)$ is discretized as an $N\times N$ complex
array and its Fourier transform is computed using the discretized FFT. 
Thus the exact algorithm consists of four steps:

\begin{enumerate}
\item Compute the forward $N$-1D transform $\psi(p_x,y)$ from $\psi(x,y)$
and multiply $\psi(p_x,y)$ grid-point by grid-point by $\e^{-\tau
  C_V(1)T_1-\tau C_V(2)V_2}$, where $T_1$ and $V_2$ are now understood
to be functions of $p_x$ and $y$.

\item Compute the 2D transform $\psi(x,p_y)$ from the updated $\psi(p_x,y)$
and multiply $\psi(x,p_y)$ by
$\e^{-\tau C_T(1)V_1-\tau C_T(2)T_2}$, where $V_1$ and $T_2$
are now functions of $x$ and $p_y$.

\item Compute the inverse 2D transform from the updated $\psi(x,p_y)$
back to $\psi(p_x,y)$ and multiply  $\psi(p_x,y)$ by
$\e^{-\tau C_V(1)T_1-\tau C_V(2)V_2}$.

\item Compute the backward $N$-1D transform from the updated $\psi(p_x,y)$ 
back to $\psi(x,y)$.
\end{enumerate}

\noindent Thus the algorithm can be implemented with only 
three 2D-FFT. (One 2D-transform = $2N$ 1D-transforms.)
This is only one 2D-FFT more than solving the non-rotational case.

\section {Solving the Gross-Pitaevskii equation}

Denoting now the entire rotating trap Hamiltonian (\ref{hrot}) as 
the operator 
\be
T=T_1+V_1+T_2+V_2,
\ee 
the corresponding 2D Gross-Pitaevskii
equation is 
\be
(\, T+g|\psi|^2 \,)\psi(x,y)=\mu\psi(x,y).
\ee
The condensate ground state can be projected out by imaginary time evolution:
\be
\psi_0\propto\lim_{\tau\rightarrow\infty}\psi(\tau)
=\lim_{\tau\rightarrow\infty}\e^{-\tau[T+V(\tau)]+\tau\mu}\psi(0).
\la{imgevo}
\ee
The chemical potential $\mu$ is determined by preserving the wave function's
normalization to unity. This will be taken for granted and this term will
be ignore in the following discussion.
Since $\psi(\tau)$ is time-dependent, we have explicitly indicated that
the Gross-Pitaevskii potential
\be
V(\tau)=g|\psi(\tau)|^2,
\la{gppot}
\ee
is also time-dependent.

In general, to solve (\ref{imgevo}) by factorization algorithms, 
one must apply rules of time-dependent factorization\cite{texp,chinchen02}:
{\it the time-dependent potential must be evaluated at an intermediate time
equal to the sum of time steps of all the $T$ operators to its right}.
For example, the first order algorithm 1A is
\be
\psi(\Delta\tau)=\e^{-\Delta\tau T}\e^{-\Delta\tau V(0)}\psi(0)
\la{alg1a}
\ee
and the first order algorithm 1B is
\be
\psi(\Delta\tau)=\e^{-\Delta\tau V(\Delta\tau)}\,\e^{-\Delta\tau T}\psi(0).
\la{alg1b}
\ee
While algorithm 1A is straightforward, 1B requires that the potential
be determined from the wave function to be computed. This
self-consistency condition can be solved by iterative methods
described below.

In contrast to real time propagation, the wave function in imaginary
time converges quickly to an {\it approximate\/} ground state depending
on $\Delta\tau$ and produces a $\psi(\Delta\tau)$ that differs from
$\psi(0)$ only by a normalization constant.  {\it Thus after some initial
iterations, the normalized $g|\psi(\Delta\tau)|^2$ is independent of
$\tau$ and can be replaced by $g|\psi(0)|^2$}. This replacement
can be justified only at small $\Delta\tau$ when the approximate wave function 
is close to the exact ground state and is unchanging in time. 
(For real time propagation, the wave function is always changing with time,
and one cannot justify this replacement even at small $\Delta t$.) At larger
$\Delta\tau$, the approximate ground state may not be a discrete bound
state and the algorithm may fail catastrophically. Thus if one
approximates $g|\psi(\Delta\tau)|^2$ by $g|\psi(0)|^2$ in
(\ref{alg1b}), then the algorithm is still first order, but only at
very small $\Delta\tau$.  We shall refer to this version of the
algorithm as 1B0.  

We define the second order algorithm 2A as 
\be
\psi(\Delta\tau)=\e^{-\frac12\Delta\tau V(\Delta\tau)}
                 \e^{-\Delta\tau T}
			 \e^{-\frac12 \Delta\tau V(0)}\psi(0)
\la{alg2a}
\ee
and algorithm 2B as
\be
\psi(\Delta\tau)=\e^{-\frac12 \Delta\tau T}
				 \e^{-\Delta\tau V(\Delta\tau/2)}
				 \e^{-\frac12 \Delta\tau T}\psi(0).
\la{alg2b}
\ee
Similarly, one can replace $g|\psi(\Delta\tau)|^2$ 
by $g|\psi(0)|^2$ in algorithm 2A without affecting its quadratic
convergence at very small $\Delta\tau$. We shall refer to this
version of the algorithm as 2A0. Algorithm 2B requires two executions 
of the exact algorithm (\ref{exfact2}) for similar convergence, 
which is less efficient. We therefore did not implement algorithm 2B.

Fig. 1 shows the convergence of algorithm 1A and 1B0 for the chemical
potential $\mu$. Both are very linear at small $\Delta\tau$.  The
calculation is done for $\delta=0.5$, $\Omega=0.5$, and $g=50$. This
choice corresponds to sizable anisotropy, rotation, coupling
strength and not close to any particular limit. The calculation uses 
$64^2$ grid points over a $14^2$ harmonic length square	
centered on the origin. Changing the grid size to $128^2$ only changes 
the stable results in the fifth or sixth decimal place. 
The ground state wave function is nearly converged by
$\tau=2$. The chemical potential shown is calculated at $\tau=10$.
Note that the linear convergence line for 1B0 fails abruptly at
$\Delta\tau\approx 0.15$. Since algorithm 2A0 is just running
algorithm 1A first followed by 1B0 at half the time-step size, the
convergence failure of 1B0 accounts for the failure of algorithm 2A0
near $\Delta\tau\approx 0.3$.  Both algorithm 1A and 1B0 require an
exceedingly small $\Delta\tau$ ($<0.001$) to produce an accurate value
of $\mu$. Even algorithm 2A0 requires $\Delta\tau$ to be $\approx
0.05$.

\section {Self-Consistent iterations}

To see the full effect of time-dependent factorization, we must 
implement 1B and 2A in the form (\ref{alg1b}) and (\ref{alg2a})
with self-consistent iterations to determined the GP potential.
The required consistency equation is of the form 
\be
\psi=\frac1{\sqrt{Z}}\e^{-\frac12 b|\psi|^2}\phi,
\la{selfcon}
\ee
where $b=c\Delta\tau g$ for some coefficient $c$,
$\phi$ is the unnormalized intermediate wave function prior to
the evaluation of the potential term, and $Z$ is 
the constant that normalizes $\psi$. This is needed because
we are solving for the GP potential, which requires 
a normalized wave function. Since only the square of the 
modulus is needed, we solve 
(\ref{selfcon}) as
\be
{x}=\frac1{Z}\e^{-b x}{a}
\la{modulus}
\ee
where $x=|\psi({\bf r})|^2$, $a=|\phi({\bf r})|^2$ and
\be
Z={\int \e^{-b x({\bf r})} a({\bf r})\, d{\bf r}}
\approx\sum_i\e^{-b x_i} a_i\,(\Delta x)^2.
\la{normz}
\ee
It is helpful to view these equations in the discrete
forms in which they are actually solved. When necessary,
we will denote array elements explicitly as 
$x_i=|\psi({\bf r}_i)|^2$, etc..

A naive way of solving (\ref{modulus})
is just to iterate,
\be
x_{n+1}=F(x_n)=\frac1{Z(x_n)}\e^{-b x_n}a.
\ee
Starting with $x_0=0$, the first iteration would 
produces the normalized $\tilde a=a/Z$,
which is a reasonable starting guess.  
Denoting $x_n=x^*+\ep_n$, where
$x^*$ is the exact solution, then
\ba
\ep_{n+1}&=&F^\prime(x^*)\,\ep_n +O(\ep_n^2)\nn\\
&=&-b x^*\left(
1-O(1/N)\right)
     \ep_n+\cdots\nn\\
&\approx&-(c\Delta\tau g x^*)\ep_n.
\la{firsterr}
\ea
The $O(1/N)$ term neglected above is from differentiating $1/Z$
with respect to $x_i$ ($=
\e^{-b\,(x_n)_i} a_i/{\sum_{j=1}^N\e^{-b\,(x_n)_j} a_j}$),
where $N$ is the total number of array elements.
It is therefore an excellent approximation to regard $Z$ as 
a constant when differentiating with respect to individual array 
elements. The error propagation (\ref{firsterr}) explains why
this naive iteration is undesirable; it 
will diverge abruptly at large $\Delta\tau$ such that $|c\Delta\tau g x^*|>1$. 

The self-consistency equation can be solved by better methods, 
such as Newton's or Halley's iterations. However, at
large $\Delta\tau$, the normalized $\tilde a$ is not a good
enough starting point. Fortunately, (\ref{modulus})  
has the exact solution
\be
x=\frac1{b}W\left(\frac{b\,a}{Z}\right)
\la{exsol}
\ee
where $W(y)$ is the Lambert W-function\cite{wfct}
\be
W\e^W=y
\ee
with series expansion
\be
W(y)=y-y^2+\frac32 y^3-\frac83 y^4 +O(y^5).
\la{expand}
\ee
The series expansion (\ref{expand}) is not useful for our purpose 
since its radius of convergence is only $1/e$. 
A better choice is the following uniform approximation by 
Winitzki\cite{wapprox}, 
\be
W(y)\approx \ln(1+y)\left(1-\frac{\ln[1+\ln(1+y)]}{2+\ln(1+x)}\right),
\la{asym}
\ee
which has a maximum relative error of 2\% (at $x \approx 2$) in [0,$\infty$].

The normalization constant $Z$ can in principle be determined from
\be
b=\int W\left(\frac{b\,a}{Z}\right) d {\bf r},
\la{fixz} 
\ee
but this integral equation is time consuming to  
solve when $W(y)$ itself has to be computed numerically.
A more workable scheme is compute compute $Z$ via (\ref{normz})
from $x_n$, and compute $x_{n+1}$ via (\ref{exsol}). The use of
(\ref{exsol}) will guarantee convergence at all $b$ and $\Delta\tau$, 
provided that one can compute $W(y)$ in a simple way. 
Thus we use the normalized $\tilde a$ as the starting value, compute $Z$ 
via (\ref{normz}), solve for $x$ via (\ref{exsol}) and normalize it to 
obtain an approximate GP potential at all $\Delta\tau$. We find that
this one iteration is sufficient to remove all instability in
algorithms 1A and 1B; further Newton-Raphson iterations produce 
marginal improvments not worth the additional effort.  
  
Fig.1 shows the convergence of algorithms 1B and 2A when the 
self-consistency condition is approximately satisfied by our
one W-function iteration. The results are denoted as 1BW and
2AW. The instability in 1B0 and 2A0 no longer appears. The
convergence of both are linear and quadratic out to large values
of $\Delta\tau$.  

For first and second order algorithms, self-consistent
iterations are not needed because $\Delta\tau$ has to be very small 
in order for these algorithms to produce results close to the
exact one. If $\Delta\tau$ is small, then one may as well use 1B0 and 2A0
without wasting time on self-consistent iterations.
Self-consistency is a concern only when one is interested in enlarging 
the step-size convergence of higher order algorithms.

\section {Forward fourth-order algorithms}

It is well known from studies of 
symplectic integrators that factorizations of the form
(\ref{alg1a}) to (\ref{alg2b}) can be generalized to 
higher order in the 
form\cite{forest,creutz,yoshida,yoshi,mcl95,suzuki,mcl02,hairer}
\be
{\rm e}^{-\Delta\tau (T+V )}=\prod_i
{\rm e}^{-a_i\Delta\tau T}{\rm e}^{-b_i\Delta\tau V},
\label{arb}
\ee
with coefficients $\{a_i, b_i\}$ determined by the required order of
accuracy.  However, as first proved by Sheng\cite{sheng} and made
explicit by Suzuki\cite{suzukinogo}, beyond second order, any
factorization of the form (\ref{arb}) {\it must} contain some negative
coefficients in the set $\{a_i, b_i\}$. Goldman and
Kaper\cite{goldman} later proved that any factorization of the form
(\ref{arb}) must contain at least one negative coefficient for {\it
both} operators.  Since a negative time step for the kinetic energy
operator will result in an unbound and unnormalizable wave function,
no such factorization scheme can be used to evolve the imaginary time
Schr\"odinger equation, including the Gross-Pitaevskii equation. To go
beyond second order, one must use {\it forward} factorization schemes
with only positive factorization
coefficients\cite{suzfour,chin97,chinchen02}. These forward algorithms
are currently the only fourth order algorithms possible for solving
time-irreversible equations with a diffusion kernel\cite{lang4,dmc4}
and have been applied successfully in solving the imaginary time
Schr\"odinger equation\cite{auer,cift03}.
Omelyand, Mryglod and Folk\cite{ome02,ome03} 
have compiled an extensive list of fourth and higher order symplectic 
algorithms. However, their sixth- and eight-order algorithms contain
negative time steps and are not forward algorithms. Recently, one of 
us have proved\cite{chinpos} that while sixth order forward algorithms 
are possible, they require an additional commutator currently not 
implementable. Thus forward algorithms are very unique. Here, we will 
show that they also yield highly accurate fourth order algorithms for 
solving the Gross-Pitaevskii equation. 

The problem we seek to solve is the ground state of 
\be
H=H_x+H_y+ V(x,y,\tau)
\la{hxyv}
\ee
where $V(x,y,\tau)$ is the GP potential (\ref{gppot}) and
\ba
H_x&=&\frac12p_y^2+\frac12(1+\delta)x^2-\Omega x p_y\,,\la{hx}\\
H_y&=&\frac12p_x^2+\frac12(1-\delta)y^2+\Omega y p_x\,.
\la{hy}
\ea
The Hamiltonian fundamentally
has three operators, which are diagonal in 
$(x,y)$, $(x,p_y)$ and $(p_x,y)$. If the external trapping 
potential $V_{ext}(x,y)$ is more
general and non-harmonic, we can still write,
\be
H=H_x+H_y+V(x,y,\tau)
\la{hxyv2}
\ee
but now with
\be
V(x,y,\tau) =V_{ext}(x,y)-\frac12(1+\delta)x^2 
-\frac12(1-\delta)y^2+g|\psi(x,y,\tau)|^2.
\la{genpot}
\ee
The parameter $\delta$ is then a free parameter associated with
algorithm, which we can choose to match the asymmetry of $V_{ext}$, or 
just set to zero. The crucial points is that, for a rotating trap, 
harmonic or not, the Hamiltonian has three operators diagonal in three 
separate spaces. By computing the density matrix of 
\be
T=H_x+H_y
\la{twoop}
\ee
exactly via algorithm (\ref{exfact2}), 
we have reduced the Hamiltonian
to a two-operator problem. This is a tremendous simplification.
This simplification is not restricted to harmonic traps,
but holds equally for an arbitrary external potential. The key point is that
the {\it rotating} part of the Hamiltonian can be diagonalized regardless 
of the choice of the confining potential. When we diagonalize $H_x+H_y$, 
we {\it generate} an inverted harmonic potential in (\ref{genpot}), 
which must be compensated by the external potential or the GP potential. 
In the following we will present results only for (\ref{hxyv}), but our algorithm 
works in the general case of (\ref{genpot}). We will come back to this point
when we discuss over-critical rotation in the next section.  

Because implementing $T$ is 
computational demanding, we must choose a fourth order algorithm with 
a minimal number of $T$ operators. Thus among the many forward algorithms
discovered so far\cite{chin97,chinchen02,chinchen03,ome02,ome03}, we 
choose to implement only the simplest algorithm, 4A.
\begin{equation}
\psi(\Delta\tau) =
\e^{ -\frac{1}{6}\Delta\tau V(\Delta\tau)} \e^{-\frac{1}{2}\Delta\tau {T}} \
\e^{-\frac{2}{3}\Delta\tau \widetilde V(\Delta\tau/2)} \
\e^{-\frac{1}{2}\Delta\tau {T}} \e^{-\frac{1}{6}\Delta\tau {V(0)}}\psi(0) \ ,
\label{alg4a}
\end{equation}
with $\widetilde V$ given by
\begin{equation}
\widetilde V=V+\frac{\Delta\tau^2}{48} 
                 \left[V,[T,V]\right].
\label{Vtilde}
\end{equation}
Despite the seeming complexity of $T$ as defined by the Hamiltonian
(\ref{hrot}), we have remarkably,
\be
[V,[T,V]]=\left (\frac{\partial V}{\partial x}\right)^2
+\left (\frac{\partial V}{\partial y}\right)^2.
\ee 
Thus the midpoint effective potential is
\be
\widetilde V(\Delta\tau/2)=g|\psi(\Delta\tau/2)|^2
+\frac{\Delta\tau^2g^2}{48}\left[
\Bigl(\frac{\partial\, |\psi(\dt/2)|^2}{\partial x}\Bigr)^2
+\Bigl(\frac{\partial\, |\psi(\dt/2)|^2}{\partial y}\Bigr)^2
\right].
 \ee
(For the more general case, $V$ is given by (\ref{genpot}))
The partial derivatives can be computed numerically\cite{auer}
by use of finite differences or FFT. Since the FFT derivative converges
exponentially with grid size, the use of FFT derivative is preferable
when the system can be made periodic. In the case with
bound state wave functions, this can be done by extending 
the grid size so the the wave function is essentially zero near
the grid edge.

To implement this fourth order algorithm, we first replace
$\psi(\dt/2)$ and $\psi(\dt)$ by $\psi(0)$. We will refer to this as
algorithm 4A00. Its convergence is shown in Fig. 2. We have retained
some first and second order results for comparison. Aside from its
abrupt instability at $\Delta\tau\approx 0.3$, its convergence is
remarkably flat. All the results at $\Delta\tau<0.3$ differ only in
the fifth decimal place.

We can also improve the convergence by making the final wave
function $\psi(\Delta\tau)$ consistent with
$V(\Delta\tau)$. The results for iterating the W-function once as
described previously is denoted as 4A0W. By just iterating the
W-function once, we extended the convergence out to $\Delta\tau\approx 0.5$.
Algorithm 4A00 and 4A0W can achieve the result of 2A0 at 
$\Delta\tau$ nearly 30 to 40 times times as large. 

To fully implement the time-dependent factorization scheme (\ref{alg4a})
and to remove the instability in 4A0W, we evolve the 
midpoint wave function $\psi(\Delta\tau/2)$ from $\psi(0)$ by a second 
order algorithm 2AW and iterate the final wave function $\psi(\Delta\tau)$ 
for consistency. We denote this algorithm as 4AWW.
Its convergence is now smooth, stable and fourth order as shown in Fig.2. 

As pointed out in Ref. \onlinecite{auer}, when the eigenfunction converges as
\be
\psi=\psi_0+O(\Delta\tau^n),
\la{wfcong}
\ee
the eigenvalue converges as
\be
E=E_0+ O(\Delta\tau^{2n}).
\la{econg}
\ee
In Fig. 3, we compare the convergence of the GP ground state energy 
\be
E=\int \psi^*(\, T+\frac12 g|\psi|^2 \,)\psi d^2 {\bf r}/\int |\psi|^2 d^2 {\bf r}.
\ee
All the fitted lines are of the monomial form
\be
E-E_0=C\Delta\tau^n.
\ee  
Algorithms 1A and 1B0 yielded near-identical quadratic
convergence. Algorithm 2AW can be fitted with a fourth order
monomial as shown. The fit is not perfect because our
W-function is only an approximation. Algorithm 4A00 and 4A0W failed 
too abruptly to show a smooth trend, but 4AWW can indeed be fitted 
with an eighth order monomial. 

The computational effort required by each algorithm is essentially
that of evaluating the exact algorithm (\ref{exfact2}), which uses 3
2D-FFT. Since 1A, 1B0, and 2A0 all use the exact algorithm once, the
second order algorithm 2A0 is clearly superior. Algorithms 4A00
requires two evaluations of the exact algorithm plus the gradient
potential. The gradient potential, if done by FFT, requires 2
2D-FFT. Thus algorithm 4A0n requires 8 2D-FFT, which is $8/3\approx3$
times the effort of algorithm 2A0. Since algorithm 4A00 converges much
better than 2A0 at time-steps more than three times as large, the
class of 4A00 and 4A0W algorithm is clearly more efficient. This efficiency is
especially evident if higher accuracy is required.  The fully
implemented algorithms 4AWW use the second order algorithm to evaluate
midpoint wave function and is therefore $\approx 4$ time the effort of
2A0. Looking at Fig. 2, algorithms 4AWW clearly converge better than
2AW (2A0) even at time-steps four times as large. Note that the first and
second order algorithms are basically similar, whereas all fourth
order algorithm are qualitatively distinct.  The second-order
algorithm is not an order-of-magnitude better than a first-order
algorithm, whereas all fourth-order algorithms are an
order-of-magnitudes better than the second-order algorithm.

This advantage of fourth order algorithms is cumulative. For
example, one can quickly evolve into the ground state by use of large
time steps.  As stated earlier, the GP ground state can be obtained at
$\tau=2$. Using algorithm 4A0W at $\Delta\tau=0.5$, one can get there
in four iterations.  Algorithm 2A0 would have taken 80 iterations at
$\Delta\tau\approx 0.02\,.$

To see how these comparisons work out in practice, we give below some
timing information. Since running time is code and machine dependent,
this should be viewed as merely illustrative. The algorithms were 
programmed in Fortran 90 and ran on a 1.2 GH Pentium machine 
with 0.5 GB of RAM. The IMSL 2D-FFT is used. For the case of a 64$\times$64
point mesh. 
each run of algorithms 1A, 1B0, 2A0, 1BW, and 2AW required 
0.0130, 0.0135, 0.0138, 0.0173 and 0.0203 second, respectively. 
(Timing is obtained by averaging over 500 run of each algorithm.)
Algorithms 1A, 1B0, 2A0 are indeed comparable, but each 
self-consistency iteration increases the time by $\approx 0.006$ second,
which is an increase of 40\% for 2AW.   
The time for algorithm 4A00,
4A0W and 4AWW are respectively 0.0394, 0.0459, 0.0630 second. 
The time for algorithm 4A00 is approximately three times that of 
2A0 and the self-consistency iteration now accounts for only 
an $16\%$ increase. Algorithm 4AWW is slightly more than four 
times that of 2A0. In the case of a 128$\times$128 point mesh, the timing for
2A0, 4A00, 4A0W, 4AWW are respectively 0.0568, 0.1669, 0.1916 and 0.2617 second.
Algorithm 4A00 is accurately three times that of 2A0 and 4AWW is
$\approx 1.5$ times that of 4A00. Everything scaled up approximately
by a factor of four.

By solving the density matrix of the rotation harmonic
oscillator (\ref{twoop}) exactly, we have effected a tremendous 
simplification and has allowed us to derive very compact fourth-order 
algorithms with excellent large time-step convergence. 
They are no more difficult to implement
than second order algorithms. If we do not have 
the exact density matrix, then we would have to approximate each
occurrence of $\e^{-\frac{1}{2}\Delta\tau {T}}$ in (\ref{alg4a}) to
fourth order, resulting in a much more complex algorithm. 

However, by solving the rotating harmonic oscillator exactly, the
current algorithms also inherited its limitations. As alluded to in Section
III, the rotating harmonic trap becomes unstable at
$\Omega_c=\sqrt{1-\delta}$. Thus if we are to use the
exact algorithm (\ref{exfact2}), we must require
$\Omega<\Omega_c$. However, it is known\cite{recati,rosen} that for $\delta<1/5$ 
and $g$ sufficiently large, the full GP equation can support
``over-critical" rotation in the interval $\sqrt{1-\delta}<\Omega<\sqrt{1+\delta}$.
For over-critical rotation, one must not prematurely impose the limitation
of the rotating harmonic oscillator on the algorithm. 

In light of our previous discussion, we now consider the case of 
\be
V_{ext}(x,y)=\frac12(1+\Delta)x^2 + \frac12(1-\Delta)y^2,
\ee
and group the Hamiltonian as follow
\be
H=H_x+H_y+V(x,y,\tau)
\la{hhxyv}
\ee
where $H_x$ and $H_y$ are defined as before in (\ref{hx}), (\ref{hy}), and 
\be
V(x,y,\tau) =\frac12(\Delta-\delta)x^2 
            -\frac12(\Delta-\delta)y^2+g|\psi(x,y,\tau)|^2.
\la{ggenpot}
\ee
We thus divorce the deformation parameter $\delta$ associated with
the algorithm, from the physical deformation parameter $\Delta$
associated with the trapping potential. If we choose $\delta=0$, the algorithm is
stable up to $\Omega=1$, above the physical critical value of $\Omega_c=\sqrt{1-\Delta}$.
Moreover, since in the Thomas-Fermi approximation the density profile follows the
shape of the potential, (\ref{ggenpot}) indeed suggests that the inverted harmonic 
potential $-\frac12\Delta y^2$ can be compensated by the GP potential at sufficiently
large $g$, making over-critical rotation possible.  

\section{Concluding summary}

In this work we have derived a number of fourth order algorithms and
demonstrate their fourth order convergence in solving the GP equation
in a rotating, anisotropic harmonic trap. These fourth order algorithms,
based on forward factorization schemes, are the only class
of factorization algorithms possible for solving evolution equations
with a diffusion kernel. Our use of the time-dependent factorization rule
provided a systematic way of solving the {\it nonlinear} GP equations
and can be generalized to solve similar nonlinear equation such as 
the Hartree-Fock and the Kohn-Sham equation\cite{aich1,aich2}. 
These fourth order algorithms 
are particularly efficient in solving for the ground state by use of large
time steps.	In constrast to other algorithms, generalizing these algorithms
to 3D is very transparent, one simply replaces 2D-FFT everywhere by 3D-FFT.  
Our use of the exact algorithm, which diagonalizes the rotating component
of the Hamiltonian, is general and can applied to any external trapping 
potential. This exact algorithm also provided insight for understanding
over-critical rotation. Physical results obtained by applying these algorithms 
will be presented elsewhere.

\begin{acknowledgments}
This work was supported in part, by a National Science Foundation 
grant (to SAC) No. DMS-0310580 and the Austrian Science Fund FWF
(to EK) under project P15083-N08).
\end{acknowledgments}
\centerline{REFERENCES}

\newpage
\begin{figure}
	\vspace{0.5truein}
	\centerline{\includegraphics[width=0.8\linewidth]{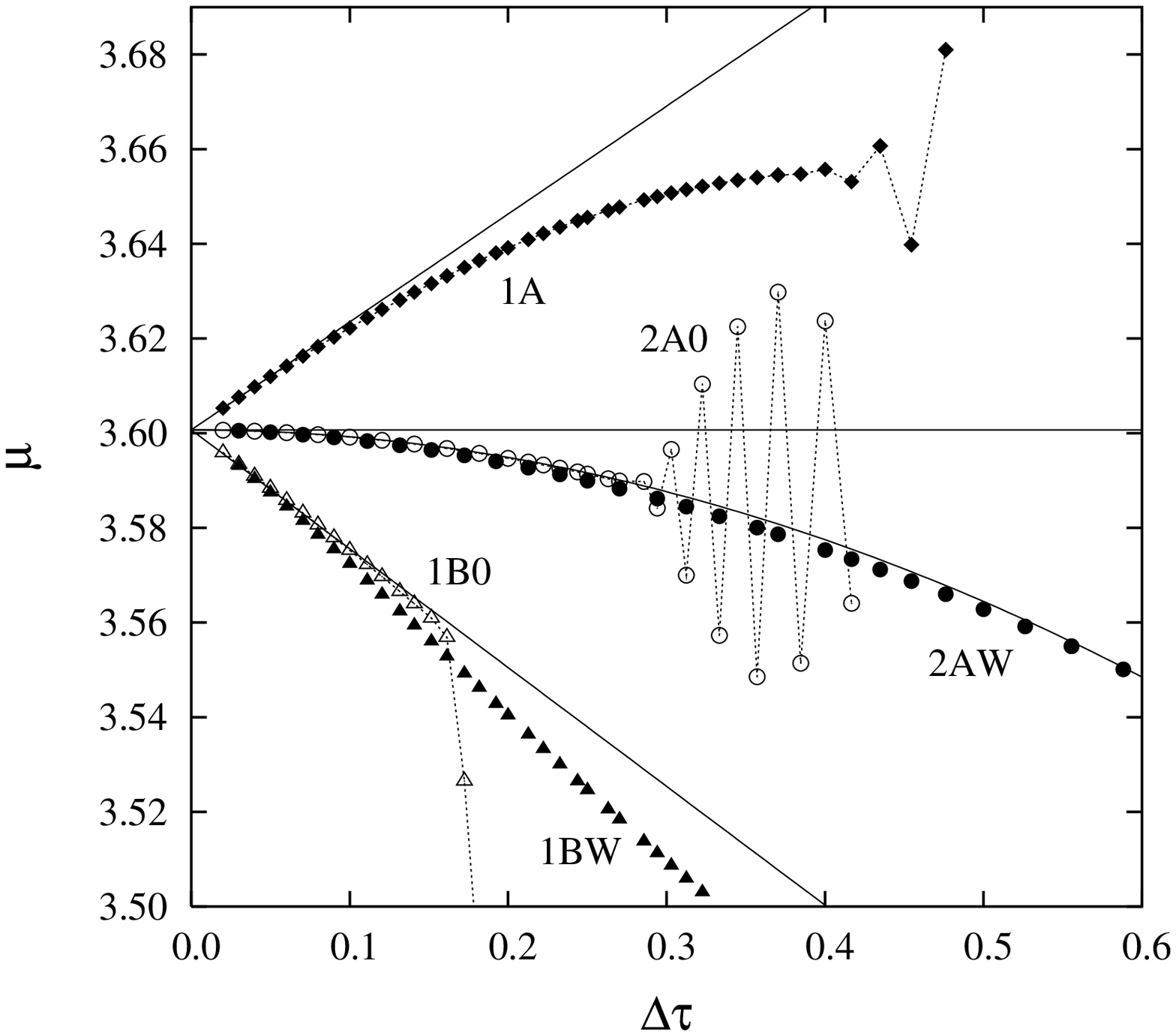}}
	\vspace{0.5truein}
\caption{Comparing the convergence of first and 
second-order algorithms in computing the chemical potential of
the Gross-Pitaevskii equation in a rotating anisotropic trap.
The lines are fitted curves to algorithm 1A, 1B0 and 2A0 to demonstrate 
the order of convergence of each algorithm. 
The instability of data points in algorithms 
1B0 and 2A0 are removed by the inclusion of one 
self-consistent W-function iteration as indicated by
1BW and 2AW.
\label{fig1}}
\end{figure}
\begin{figure}
	\vspace{0.5truein}
	\centerline{\includegraphics[width=0.8\linewidth]{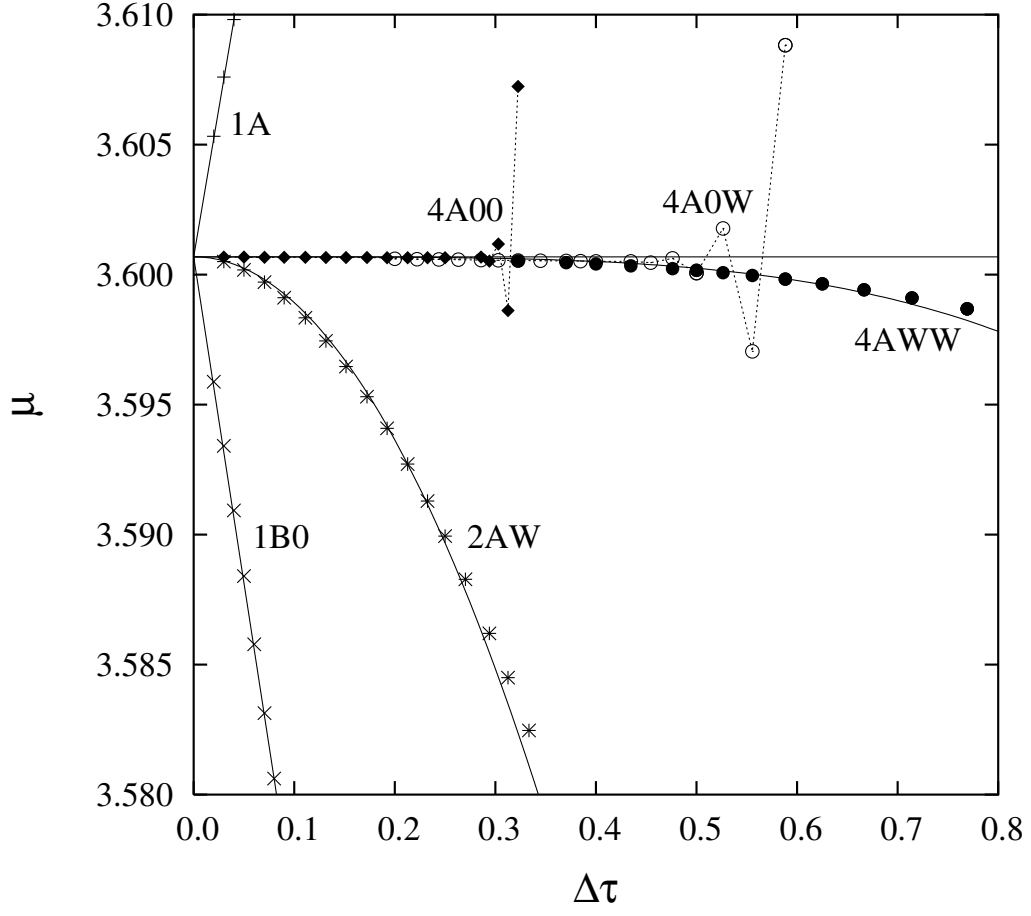}}
	\vspace{0.5truein}
\caption{The convergence of fourth-order algorithms
in computing the chemical potential of the Gross-Pitaevskii equation
in a rotating trap. Algorithms 4A00 (solid diamonds) and 4A0W (circles)
are unstable beyond $\Delta\tau\approx 0.3$ and $0.45$ respectively. 
Algorithm 4AWW (solid circles) is stable and showed excellent fourth order convergence.  
\label{fig2}}
\end{figure}
\begin{figure}
	\vspace{0.5truein}
	\centerline{\includegraphics[width=0.8\linewidth]{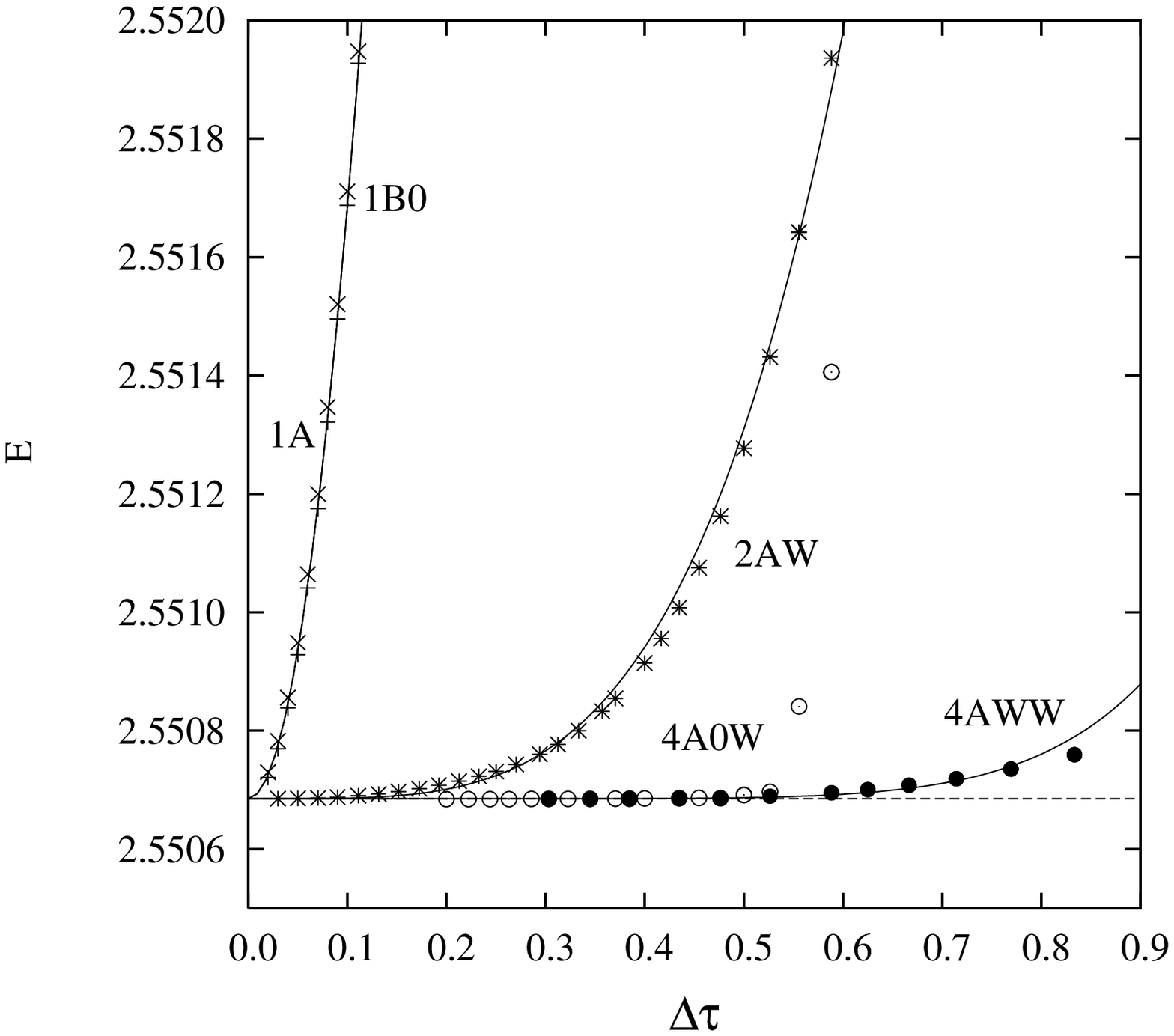}}
	\vspace{0.5truein}
\caption{The convergence of various algorithms in computing
the ground state energy of the GP equation. Both
first order results showed near-identical quadratic convergence. The second order
result 2AW (asterisks) is fourth order and 4AWW (solid circles)
is eighth order. 
Results for 4A0W (circles) cannot be fitted because instability sets in abruptly at 
$\dt\approx 0.5$. Results for 4A00 is similar with instability at $\approx 0.3$
and is not shown.   
\label{fig3}}
\end{figure}


\begin{thebibliography}{}
\bibitem{dalf}F. Dalfovo, S. Giorgini, L. Pitaevskii and S. Stringari,
              Rev. Mod. Phys. {\bf 71}, 463 (1999).
\bibitem{fet}A. Fetter and A. Svidzinsky, J. Phys.: Condens. Matter 
                 {\bf 13}, R135 (2001).
\bibitem{fetter}A. Fetter, B. Jackson and S. Stringari, Phys. Rev.
                 {\bf A71}, 013605 (2005).
\bibitem{taha}T. R. Taha and M. J. Ablowitz, J. Comput. Phys. {\bf 55}, 203 (1984).
\bibitem{jackson}B. Jackson, J. F. McCann and C. S. Adams, J. Phys. {\bf B 31}, 4489 (1998).
\bibitem{dion} C. M. Dion and E. Cances, Phys. Rev. {\bf E 67}, 046706 (2003).
\bibitem{adhi} S. K. Adhikari and P. Muruganandam, J. Phys. {\bf B 35}, 2831 (2002)
\bibitem{rupr} P. A. Ruprecht, M. J. Holland, K. Burnett, and M. Edwards, 
          Phys. Rev. {\bf A 51}, 4704 (1995).
\bibitem{band}A. D. Bandrauk and H. Shen, J. Phys. {\bf A 27}, 7147 (1994).
\bibitem{lang4}H. A. Forbert and S. A. Chin,
                Phys. Rev. {\bf E 63}, 016703 (2001).
\bibitem{dmc4}H. A. Forbert and S. A. Chin,
                Phys. Rev. {\bf B 63}, 144518 (2001).
\bibitem{auer}J. Auer, E. Krotscheck, and S. A. Chin,
                J. Chem. Phys. {\bf 115}, 6841 (2001).
\bibitem{cift03}O. Ciftja and S. A. Chin, Phys. Rev.
               {\bf B 68}, 134510 (2003).
 \bibitem{suzfour}M. Suzuki, {\it Computer Simulation Studies in 
            Condensed Matter Physics VIII},
           eds, D. Landau, K. Mon and H. Shuttler (Springler, Berlin, 1996).
\bibitem{chin97} S. A. Chin, Phys. Lett. {\bf A226}, 344 (1997).
\bibitem{chinchen02}S. A. Chin and C. R. Chen,
                J. Chem. Phys. {\bf 117}, 1409 (2002).
\bibitem{chinchen03}S. A. Chin, and C. R. Chen,
         "Forward Symplectic Integrators for Solving Gravitational
          Few-Body Problems", arXiv, astro-ph/0304223, in press,
          Cele. Mech. and Dyn. Astron.
\bibitem{ome02}I. P. Omelyan, I. M. Mryglod and R. Folk,
               Phys. Rev. {\bf E66}, 026701 (2002).
\bibitem{ome03}I. P. Omelyan, I. M. Mryglod and R. Folk,
               Comput. Phys. Commun. {\bf 151} 272 (2003)
\bibitem{edwards} M. Edwards {\it et al.}, Phys. Rev. {\bf A53}, R1950 (1996).
\bibitem{fey}R. P. Feynman, {\it Statistical Mechanics -A set of
             Lectures}, Benjamin Advanced Book, Reading, MA, 1972. 
\bibitem{ring} P. Ring and P. Schuck, ``The Nuclear Many-Body Problem",
        P.133, Springer-Verlag, Berlin-NY (1980).
\bibitem{ok} M.\"O. Oktel, Phys. Rev. {\bf A69}, 023618 (2004). 
\bibitem{texp}M. Suzuki, Proc. Japan Acad. {\bf 69}, Ser. B, 161 (1993).
\bibitem{wfct} R. M. Corless, G. H. Gonnet, D. E. G. Hare, D. J. Jeffrey 
               and D.E. Knuth, Adv. Comput. Math {\bf 5}, 329 (1996). 
\bibitem{wapprox} S. Winitzki, {\it Lecture Notes in Computer Science}, 
              Springer-Verlag, {\bf 2667}, 780 (2003).
\bibitem{forest}E. Forest and R. D. Ruth, Physica D {\bf 43}, 105 (1990).
\bibitem{creutz}M. Creutz and A. Gocksch, Phys. Rev. Letts. {\bf 63}, 9 (1989).
\bibitem{yoshida}H. Yoshida, Phys. Lett. {\bf A150}, 262 (1990).
\bibitem{yoshi} H. Yoshida, Celest. Mech. {\bf 56} (1993) 27.
\bibitem{mcl95} R. I. McLachlan, SIAM J. Sci. Comput. {\bf 16}, 151 (1995).
\bibitem{suzuki}M. Suzuki, Phys. Lett. {\bf A146}, 319 (1990); {\bf 165}, 
                387 (1992).
\bibitem{mcl02} R. I. McLachlan and G. R. W. Quispel, Acta Numerica,
               {\bf 11}, 241 (2002).
\bibitem{hairer}{\it Geometric Numerical Integration}, by E. Hairer,
                C. Lubich, and G. Wanner, 
                Springer-Verlag, Berlin-New York, 2002.
\bibitem{sheng}Q. Sheng, IMA Journal of numberical anaysis, {\bf 9}, 
              199 (1989). 
\bibitem{suzukinogo}M. Suzuki, J. Math. Phys. {\bf 32}, 400 (1991).
\bibitem{goldman}D. Goldman and T. J. Kaper, SIAM J. Numer. Anal.,{ \bf 33}, 
                   349 (1996).
\bibitem{chinpos} S. A. Chin, Phys. Rev. E {\bf 71}, 016703 (2005).
\bibitem{recati}A. Recati, F. Zambelli and S. Stringari, 
                Phys. Rev. Letts. {\bf 86}, 377 (2001).
\bibitem{rosen}P. Rosenbusch {\it et al.}, 
               Phys. Rev. Letts. {\bf 88}, 250403 (2002).
\bibitem{aich1} M. Aichinger, S. A. Chin, and E. Krotscheck,
               Computer Physics Communications, in press
\bibitem{aich2} M. Aichinger, S. A. Chin, E. Krotscheck, 
                and E. R\"as\"anen, Phys. Rev. B. (submitted)
\end{thebibliography}

\end{document}